\documentclass[aps,prd,twocolumn,superscriptaddress,nofootinbib]{revtex4}
\usepackage{graphicx}
\usepackage{bm}
\usepackage{tabularx}
\usepackage{epsfig}
\usepackage{amsfonts}
\usepackage{amsmath}
\usepackage{ upgreek }
\usepackage{color}
\usepackage{color}

\newcommand{\la}{\label}
\newcommand{\bbm}{\begin{multline}}
\newcommand{\eem}{\end{multline}}
\newcommand{\be}{\begin{equation}}
\newcommand{\ee}{\end{equation}}
\newcommand{\bea}{\begin{eqnarray}}
\newcommand{\eea}{\end{eqnarray}}
\newcommand{\p}{\partial}

\newcommand{\comment}[1]{}

\begin{document}

%%%%%%%%%%%%%%%%%%%%%%%%%%%%
%%%%%%%%%%%%%%%%%%%%%%%%%%%%
%%%%%%%%%%%%%%%%%%%%%%%%%%%%

\title{\large{Hydrodynamics with gauge anomaly: \\
Variational principle and Hamiltonian formulation}}

%%%%%%%%%%%%%%%%%%%%%%%%%%%%
\author{Gustavo M.~Monteiro}
\affiliation{Department of Physics and Astronomy, Stony Brook University, Stony Brook, NY 11794, USA}

\author{Alexander G.~Abanov}
\affiliation{Department of Physics and Astronomy, Stony Brook University,  Stony Brook, NY 11794, USA}
\affiliation{Simons Center for Geometry and Physics,
Stony Brook University,  Stony Brook, NY 11794, USA}

\author{V.~P.~Nair}
\affiliation{Physics Department, City College of the CUNY, New York, NY 10031, USA}
%%%%%%%%%%%%%%%%%%%%%%%%%%%%

\begin{abstract}
We present a variational principle for relativistic hydrodynamics with gauge-anomaly terms for a fluid coupled to an Abelian background gauge field. For this we utilize the Clebsch parametrization of the velocity field. We also set up the Hamiltonian formulation and the canonical framework for the theory. While the equations of motion only involve the density and velocity fields, i.e., the Clebsch potentials only appear in the combination which is the velocity field, the generators of symmetry transformations (including the Hamiltonian) depend explicitly on one of the Clebsch potentials, if the background field is time-dependent. For the special case of time-independent background fields, this feature is absent.
\bigskip
\bigskip
\end{abstract}

\maketitle

%%%%%%%%%%%%%%%%%%%%%
\section{\protect Introduction}
%%%%%%%%%%%%%%%%%%%%%

Hydrodynamics is a long-wavelength effective description of interacting systems based on the assumption of local equilibrium. Hydrodynamic equations are essentially local conservation laws supplemented by the constitutive relations between conserved densities. These conservation laws are macroscopic manifestations of symmetries of the system. Constitutive relations are often written phenomenologically and involve unknown ``equations of state'', which in principle should be obtainable from the underlying ``microscopic'' theory such as kinetic theory, many body models or quantum field theory \cite{LandauLifshitz-6}.

If the underlying theory is a quantum field theory (QFT) with \emph{quantum anomalies}, the conservation laws corresponding to anomalous symmetries are broken. However, the anomalous symmetry breaking is rather subtle and one might hope for an applicability of a universal hydrodynamic description with additional hydrodynamic terms taking anomalies into account.  This possibility was noticed initially in AdS/CFT systems \cite{2008-ErdmengerHaackKaminskiYarom, 2008-BanerjeeEtAl}, and then in genuine relativistic hydrodynamic formulation by Son and Surowka for a particular case of Abelian gauge anomaly \cite{2009-SonSurowka} .

The goal of this work is to find a variational and Hamiltonian formulations of the hydrodynamics with gauge anomaly \cite{2009-SonSurowka}. Variational and Hamiltonian approaches to hydrodynamics have a long history and we refer the reader to Refs.~\onlinecite{1997-ZakharovKuznetsov},
\onlinecite{2004-JNPPreview} for reviews.

Let us start with equations of anomalous hydrodynamics of \cite{2009-SonSurowka}.
The current and energy-momentum conservation laws for anomalous QFT in the background gauge field can be written as:
\bea
	\partial_\lambda j^\lambda 
	&=& -\frac{C}{8} \epsilon^{\lambda\nu\sigma\tau}F_{\lambda\nu}F_{\sigma\tau} \,,
 \la{dJC} \\
	\partial_\lambda  T^{\lambda\nu}
	&= & F^{\nu\sigma} j_\sigma \,.
 \la{dTC} 
\eea
The right hand side of the equation (\ref{dTC}) is the Lorentz force, while the right hand side of (\ref{dJC}) is the gauge anomaly term, fully characterized by a single dimensionless constant $C$. Here and in the following we drop the angular brackets denoting expectation values, e.g., $\langle j\rangle\to j$, so that $j^{\lambda}$ and $T^{\lambda\nu}$ are classical fields representing the
current and the energy-momentum tensor. 

Assuming local equilibrium and imposing the local form of the second law of thermodynamics Son and Surowka were able to constrain the form of constitutive relations to \cite{2009-SonSurowka}:
\bea
	j^\lambda &=& nu^\lambda+\frac{C}{12}
	\epsilon^{\lambda\nu\sigma\tau}\,
	\mu u_\nu\left(2\mu\,\partial_\sigma u_\tau+3F_{\sigma\tau}\right) \,,
 \label{JuC} \\
	T^{\lambda\nu} &=& n\mu\, u^\lambda u^\nu+P(\mu) \,g^{\lambda\nu} \, .
 \label{TuC} 
\eea
Here we have introduced the equation of state of the fluid $P(\mu)$ which gives the fluid pressure $P$ as a function of the chemical potential $\mu$. The charge density in the fluid rest frame is given by $n = P'(\mu)$. The fluid 4-velocity $u^\lambda$ satisfies $u^\lambda u_\lambda=-1$ and, therefore, has only three independent components.

In this paper we are interested in the case of zero temperature. The constitutive relations (\ref{JuC},\ref{TuC}) are the specifications of more general relations of \cite{2009-SonSurowka} to the case of zero temperature and the absence of dissipation. In this case, the zeroth component of the equation (\ref{dTC}) --- the energy conservation --- is not independent, but can be viewed as a consequence of the other four equations (\ref{dJC},\ref{dTC}). The latter four independent equations fully determine the evolution of $n$ and three independent components of 4-velocity $u^\lambda$.

We notice that equations (\ref{dJC}-\ref{TuC}) constitute the first-order hydrodynamics equations written in Landau frame. Namely, the constitutive relations (\ref{JuC},\ref{TuC}) are first order in derivatives and the ambiguity in the definition of 4-velocity is resolved by defining it as an eigenvector of the energy-momentum tensor. Landau frame was used in \cite{LandauLifshitz-6} and was adopted in \cite{2009-SonSurowka} to construct the hydrodynamics with gauge anomaly.

Attempts to find a variational principle for equations (\ref{dJC}) and (\ref{dTC}) have had
only partial success so far \cite{2013-HaehlLoganagayamRangamani}. \footnote{ For the one-dimensional case check \cite{2011-DubovskyHuiNicolis}. } These approaches rely on an effective action for the Lagrangian specification of fluid variables \cite{2012-DubovskyHuiNicolisSon, 2011-NicolsSon-HallViscosity}. The action principle for non-abelian hydrodynamics was presented in \cite{2004-JNPPreview}, where the authors introduced the idea of coarse graining the coadjoint orbit action. A similar approach to fluid dynamics for spinning particles has been recently developed in 
\cite{2014karabali-nair}.  An action that includes anomalies in the standard model of particle physics within the framework of the coadjoint orbit method was given in \cite{2012NairRayRoy}. 
The anomaly structure in the standard model is different from what is given in
 (\ref{dJC}-\ref{TuC}) and so the effective action for anomalies in \cite{2012NairRayRoy} is
 not immediately applicable to the present problem.
In this work, we give a variational principle that produces the Son-Surowka equations at zero temperature. This approach uses the so-called Clebsch potentials to parametrize the Eulerian variables \cite{Clebsch}. We restrict ourselves to the flat Minkowski spacetime, though the generalization to more general geometric backgrounds is straightforward. Unless otherwise specified, we use the Cartesian orthonormal frame, where the pseudo-metric can be chosen as $g_{\lambda\nu}=\mbox{diag}(-1,1,1,1)$. 

The variational principle and the symmetries are analyzed in sections II and III. 
Using the obtained action,
we then derive the corresponding Hamiltonian formulation specifying the form of the relativistic Hamiltonian and the Poisson brackets. 
We emphasize the symmetries of the system and their manifestations in Hamiltonian formalism,
pointing out the special feature of one of the Clebsch potentials
appearing separately and not via the combination in the dynamic velocity field.
This feature is commented on in section VII and we
conclude with the discussion of the obtained results and their possible generalizations. 

%%%%%%%%%%%%%%%%%%%%%
\section{Hydrodynamic Action} 
%%%%%%%%%%%%%%%%%%%%%

The variational principle for perfect relativistic fluid dynamics is well known \cite{lin, schutz1970perfect, schutz1971hamiltonian}. In the following we find an additional term in the hydrodynamic action of \cite{schutz1970perfect, schutz1971hamiltonian} reproducing the gauge anomaly in hydrodynamic equations. 

The field content of the hydrodynamic action is given by 4 components of the 4-current $J^\lambda$ and 3 scalar Clebsch potentials $(\theta,\alpha,\beta)$ parametrizing dynamic velocity $\xi_\lambda$
\bea
	\xi_\lambda = \p_\lambda\theta +\alpha\p_\lambda \beta \,.
 \la{xiClebsch}
\eea
Then one of the main results of this work is that the action generating equations (\ref{dJC}-\ref{TuC}) is given by:
\begin{eqnarray}
	S=&-&\int \left[J^\lambda\left(\xi_\lambda-A_\lambda\right)
	-\varepsilon(n)\right]\,d^4x\; + \nonumber
 \\
	&+&\frac{C}{6}\int A\wedge\xi \wedge d\left(\xi+A\right) \,. 
 \label{RelAction}
\end{eqnarray}
Here $\varepsilon(n)$ is the proper energy density of the fluid which is assumed to be a known function of the proper charge density $n$. The latter is given by an absolute value of the 4-current $J^\lambda$ as $n\equiv\sqrt{-g_{\lambda\nu} J^\lambda  J^\nu}$. The second line of (\ref{RelAction}) describes the anomaly and is written in the differential form language (e.g., $\xi \equiv \xi_\lambda dx^\lambda$ is a 1-form so that $\xi=d\theta+\alpha d\beta$ etc.). Taking $C=0$ in (\ref{RelAction}) we recover the action for a relativistic perfect fluid without anomaly  \cite{schutz1970perfect, schutz1971hamiltonian}.

The full set of variational equations is obtained by varying (\ref{RelAction}) over $J^\lambda,\theta,\alpha,\beta$.
We start with 
\begin{equation}
	\frac{\delta S}{\delta J^\lambda}
	=-\left(\xi_\lambda-A_\lambda\right)+ \varepsilon'(n)\frac{J_\lambda}{n}=0 \,.  
 \label{EoMJ}
\end{equation}
It is convenient to introduce a complete parametrization of the 4-current $J^\lambda$ in terms of its absolute value $n$ and its direction given by 4-velocity $u^\lambda$ as
\be
	J^\lambda \equiv n u^\lambda\,, \qquad u^\lambda u_\lambda =-1 \,.
 \la{Jnu}
\ee
Then equation (\ref{EoMJ}) can be viewed as a relation between the dynamic velocity, density and the 4-velocity
\be
	\xi_\lambda - A_\lambda = \mu \,u_\lambda\,,
 \la{xiu}
\ee
where the chemical potential $\mu(n)$ is given by the derivative of the energy density as
\be
	\mu(n) \equiv \varepsilon'(n)\,.
 \la{muepsilon}
\ee

The Clebsch potentials $\theta,\alpha,\beta$ enter (\ref{RelAction}) only through $\xi$ given by (\ref{xiClebsch}). The corresponding variations give the following equations of motion
\bea
	\frac{\delta S}{\delta\theta}
	&=& \partial_\lambda \left(\frac{\delta S}{\delta\xi_\lambda}\right) = 0\,, 
 \label{EoMtheta} \\
	\frac{\delta S}{\delta\alpha}
	&=& \frac{\delta S}{\delta\xi_\lambda}\partial_\lambda\beta =0 \,, 
 \label{EoMalpha} \\
	\frac{\delta S}{\delta\beta}
	&=& \partial_\lambda\left(\alpha \frac{\delta S}{\delta\xi_\lambda}\right) 
	=\frac{\delta S}{\delta\xi_\lambda}\partial_\lambda\alpha=0\,, 
 \label{EoMbeta}
\eea
with
\begin{equation}
	-\frac{\delta S}{\delta\xi_\lambda}
	=nu^\lambda+\frac{C}{6}\epsilon^{\lambda\nu\eta\sigma}
	\left[2A_{\nu}\p_{\eta}\xi_{\sigma}-(\xi_{\nu}-A_\nu)\p_{\eta}A_{\sigma}\right] \,.
 \la{Sxi}
\end{equation}
Introducing the charge current 
\begin{equation}
	j^\lambda=-\frac{\delta S}{\delta \xi_\lambda}
	+\frac{C}{6}\epsilon^{\lambda\nu\eta\sigma}\left[3\,\partial_\nu(A_\eta\xi_\sigma)-
	3\,A_\nu\partial_\eta A_\sigma+\xi_\nu\partial_\eta\xi_\sigma\right] \,,
 \la{chcurrent}
\end{equation} 
we obtain (\ref{dJC}) from (\ref{EoMtheta}) and (\ref{xiClebsch}). The relations (\ref{chcurrent},\ref{Sxi})
give the constitutive relation (\ref{JuC}).

Defining the energy-momentum tensor by (\ref{TuC}), one can derive the conservation law (\ref{dTC}) from (\ref{xiu}) and (\ref{EoMtheta}-\ref{EoMbeta}) after some tedious but straightforward manipulations.\footnote{Technical remark: it is convenient to start this derivation with an obvious equation $\frac{\delta S}{\delta\xi_\lambda}\left[\partial_\lambda\left(\frac{\delta S}{\delta J^\nu}\right)-\partial_\nu\left(\frac{\delta S}{\delta J^\lambda}\right)\right]=0$.}
We do not go through this derivation in more detail, since, in the next section
\ref{sec:sym}, we will derive equations (\ref{dJC}-\ref{TuC}) more straightforwardly from symmetries of the action (\ref{RelAction}). 

In the absence of the gauge field background $A_\mu=0$ the action (\ref{RelAction}) becomes the conventional action for relativistic perfect fluid dynamics\cite{schutz1970perfect, schutz1971hamiltonian}. The only manifestation of the gauge anomaly in this case is the non-conventional relation between current and 4-velocity. Namely, the relation (\ref{JuC}) becomes $j^\lambda = nu^\lambda +\frac{C}{3}\mu^2 \omega^\lambda$ with relativistic vorticity defined as $\omega^\lambda =\frac{1}{2}\epsilon^{\lambda\nu\sigma\tau}u_\nu \p_\sigma u_\tau$. This current is conserved $\p_\lambda j^\lambda=0$ because both relations $\p_\lambda (nu^\lambda)=0$ and $\p_\lambda (\mu^2\omega^\lambda)=0$ follow from (\ref{RelAction}) in the absence of the gauge background.\footnote{One can think of the second relation as a consequence of (\ref{dJC},\ref{dTC}). We notice that the second conserved quantity $\mu^2 \omega^\lambda$ can be identified as a density of the Casimir (helicity) of the relativistic perfect fluid dynamics.} This simple ``removal'' of anomaly by current redefinition is not possible though when a non-trivial gauge field background is present.

%%%%%%%%%%%%%%%%%%%%%
\section{Symmetries}
 \la{sec:sym}
%%%%%%%%%%%%%%%%%%%%%

In this section we show explicitly that the equations (\ref{dJC},\ref{dTC}) can be obtained as consequences of (anomalous) gauge symmetry and space-time translational symmetry of the action (\ref{RelAction}), respectively.

We notice that the first line of (\ref{RelAction}) is symmetric with respect to the gauge transformation
with the gauge parameter $\Lambda(x)$
\bea
	\delta_{\Lambda}A_\lambda = \p_\mu\Lambda\,,
	\qquad \delta_{\Lambda}\theta = \Lambda\,.
 \la{gt}
\eea
Indeed, from (\ref{xiClebsch},\ref{gt}) we have $\delta_\Lambda \xi_\lambda = \p_\lambda\Lambda$ and see that the combination $\xi_\lambda-A_\lambda$ entering (\ref{RelAction}) is gauge invariant.

This gauge invariance, however, is broken by the anomalous (second line) part of the action (\ref{RelAction}). It is easy to verify that,
up to boundary terms, the gauge transformation of the action is given by
\bea
	\delta_\Lambda S 
	&=& \int  \partial_\lambda\Lambda\left(\frac{\delta S}{\delta\xi_\lambda}
	+\frac{\delta S}{\delta A_\lambda}\right)d^4x
	 = \frac{C}{6}\int \Lambda\,dA\wedge dA\,.
 \nonumber \\
 	&&
 \la{LambdaS}
\eea
Unlike the case of a general breaking of a symmetry, the loss of symmetry due to anomalies is
rather special.  The gauge variation of the action depends only on the background gauge field and has a very specific form, the latter being determined by the densities of certain topological invariants.
It is easy to see that the action can be made fully gauge invariant by supplementing it with
 the Chern-Simons term $-\frac{C}{6}\int_{M^5} A\wedge dA\wedge dA$. The integral in this term is taken over an auxiliary 5-dimensional space $M^5$ which boundary coincides with the physical space-time. This gives an elegant interpretation of the anomaly of the 4-dimensional theory
 as being due to the inflow of charge from the fifth dimension, a
set-up known as \emph{anomaly inflow}; this is standard and well known in QFT with quantum anomalies \cite{genRefAnomalyInflow}.

With the variation with respect to the Clebsch potential $\theta$ satisfying the
(\ref{EoMtheta}), the variation of (\ref{LambdaS}) over $\Lambda$ gives the charge conservation law modulo the anomaly as
\begin{equation}
	\partial_\lambda\left(\frac{\delta S}{\delta A_\lambda}\right)
	=-\frac{C}{24} \epsilon^{\lambda\nu\sigma\tau}F_{\lambda\nu}F_{\sigma\tau}.
	\label{conserve1}
\end{equation}
The quantity $\delta S/\delta A_\lambda$ is known as the \emph{consistent current} versus the \emph{covariant current} $j^\lambda$ defined in (\ref{JuC}). A quick calculation shows that
\begin{equation}
	j^\lambda
	=\frac{\delta S}{\delta A_\lambda}
	-\frac{C}{6}\epsilon^{\lambda\nu\sigma\tau}A_\nu F_{\sigma\tau} \,. 
 \label{jcons}
\end{equation}
Taking the divergence of (\ref{jcons}), we obtain (\ref{dJC}).

We now turn to the energy-momentum conservation (\ref{TuC}). The standard way of deriving this law is to gauge space-time translational symmetries by introducing the background metric and study the invariance of the action under diffeomorphisms $x^\lambda \to x^\lambda +\zeta^\lambda(x)$.

We consider (\ref{RelAction}) in an arbitrary background metric
by replacing the measure $d^4x$ by the invariant one $\sqrt{-g}\, d^4x$ and by introducing the metric into all scalar products. Notice that $\xi_\lambda$ is naturally a covariant vector, being derivatives
of the scalar Clebsch potentials, and thus $J^\lambda \, \xi_\lambda$ being an invariant scalar product does not require additional metric factors.
However, a scalar product like $J^2$ will become
$J^\mu J^\nu \, g_{\mu\nu}$.
The resulting action is invariant under diffeomorphisms, i.e., $\delta_{\zeta}S=0$, and on equations of motion we have
\begin{equation}
	\int \left[(\mathcal L_\zeta \boldsymbol g)_{\nu\lambda}
	\frac{\delta S}{\delta g_{\nu\lambda}}
	+(\mathcal L_\zeta A)_\lambda\frac{\delta S}{\delta A_\lambda}\right]d^4x
	=0 \, ,
 \la{zetaS}
\end{equation}
since the terms corresponding to the variations of the fields vanish by the equations of motion.
Here $\mathcal L_\zeta$ denotes the Lie derivative with respect to the vector field $\zeta$.
Explicitly
\bea
	(\mathcal L_\zeta \boldsymbol g)_{\nu\lambda}
	&=& \partial_\nu\zeta_\lambda+\partial_\lambda\zeta_\nu\,,
 \\
	(\mathcal L_\zeta A)_\lambda
	&=& \zeta^\nu F_{\nu\lambda}+\partial_\lambda(\zeta^\nu A_\nu) \,.
\eea
Using these formulas and setting the coefficient of $\zeta^\nu$ 
in (\ref{zetaS})  to zero
we obtain \footnote{The identity $A_\nu\,\epsilon^{\lambda\eta\sigma\tau}F_{\lambda\eta}F_{\sigma\tau}=-4F_{\nu\lambda}\,\epsilon^{\lambda\eta\sigma\tau}A_\eta F_{\sigma\tau}$ can be useful.} 
\begin{equation}
	 \partial_\lambda T^\lambda\,_\nu
	=F_{\nu\lambda}\frac{\delta S}{\delta A_\lambda}
	-\frac{C}{6}F_{\nu\lambda}\,\epsilon^{\lambda\eta\sigma\tau}A_{\eta}F_{\sigma\tau}\, , 
 \label{EMConsAnom}
\end{equation}
with 
\be
	T^{\lambda\nu}\equiv-\frac{2}{\sqrt{-g}}\frac{\delta S}{\delta g_{\lambda\nu}}\,,
 \la{Tdef}
\ee 

A quick calculation shows that the energy-momentum tensor (\ref{Tdef}) is the same as (\ref{TuC}). This is expected as the last term of (\ref{RelAction}) is the integral of a 4-form --- which is metric-independent --- and gives no contribution to the energy-momentum tensor. Therefore, (\ref{TuC}) is identical in form to the 
energy-momentum tensor for conventional perfect fluid dynamics. We see that the metric independence of the anomalous contribution to (\ref{RelAction}) is an essential feature of the analysis in the hydrodynamic Landau frame where the energy-momentum tensor is not modified by corrections
which are of the first order in gradients of the velocity.

Finally, it is easy to see that the equation (\ref{EMConsAnom}) with the relation (\ref{jcons}) is equivalent to (\ref{dTC}). This completes the demonstration that the action (\ref{RelAction}) does indeed reproduce
equations (\ref{dJC}-\ref{TuC}).

%%%%%%%%%%%%%%%%%%%%%
\section{Hamiltonian Formalism}
%%%%%%%%%%%%%%%%%%%%%

In this section we set up the Hamiltonian formulation of equations (\ref{dJC}-\ref{TuC}) 
starting with the action (\ref{RelAction}).

We start by reducing the seven independent variational fields of (\ref{RelAction}) to four given by $J^0$ and by the Clebsch parameters $\theta,\alpha,\beta$. 
The spatial components of (\ref{Jnu},\ref{xiu}) give
\be
	J_i = \frac{n}{\mu}(\xi_i-A_i)\,
 \la{Jixi}
\ee
and we can eliminate the spatial components of the current $J^i$ using (\ref{Jixi}). 
Using this relation and the defining relation (\ref{Jnu}) for $n$, 
namely,  $(J^0)^2-(J^i)^2=n^2$, we find
\be
	J^0\equiv \rho = \frac{n}{\mu}\sqrt{\mu^2+(\xi_i-A_i)^2} \,.
 \la{rhon}
\ee
Here and in the following we use $\rho$ to denote $J^0$. We may regard
$J^0 =\rho$ as the independent variable, with $n$ given
implicitly as a function of $\rho$ by (\ref{rhon}).\footnote{As $\mu(n)$ is assumed to be a known function of $n$ (\ref{muepsilon}) the equation (\ref{rhon}) can in principle be solved to obtain $n(\rho,\xi_i)$, $\mu(\rho,\xi_i)$ etc.}

Substituting (\ref{Jixi},\ref{rhon}) into (\ref{RelAction}) we obtain the action in a form linear in 
the time-derivatives and depending only on fields $\rho,\theta,\alpha,\beta$. After some integrations by parts, it can be brought to the following form:
\be
	S=\int \left(\langle\pi_\theta\,,\dot\theta\rangle+\langle\pi_\beta\,,\dot\beta\rangle
	-H\right)dt \,, 
 \la{S-notation}
\ee
where $ \langle f, g\rangle\equiv\int f(x)g(x)\,d^3x$
denotes the $L^2$-inner product in the space of real functions,  $H$ is the Hamiltonian, $\pi_\theta$ and $\pi_\beta$ are the canonical field momenta conjugate to $\theta$ and $\beta$, respectively. The explicit formulas for the canonical momenta are:
\bea
	\pi_\theta 
	&=&-\left[\rho+\frac{C}{6}\,(A_i+\alpha\,\partial_i\beta)\, B^i\right]\,, 
 \la{pitheta} \\
	\pi_\beta
	&=& -\alpha\left[\rho+\frac{C}{6}\,(A_i-\partial_i\theta)\, B^i\right] \, .
 \la{pibeta}
\eea
The Hamiltonian $H$ in (\ref{S-notation}) is given by
\bea
	H &=&\int \left[\rho\sqrt{\mu^2+(\xi_i-A_i)^2}
	-P(\mu)-\rho A_0\right] d^3 x\; 
 \nonumber \\
	&-& \frac{C}{6}\int \left[\xi_iB^i A_0
	+\epsilon^{ijk}(\partial_i\theta-A_i)\,\xi_jE_k\right] d^3 x \,. 
 \label{H-time}
\eea
The pressure $P(\mu)$ is related to the energy density by the Legendre transform $\varepsilon(n)=n\mu-P(\mu)$, with $P'(\mu)=n$ and we have also introduced the magnetic and electric fields $B^i= \epsilon^{ikl}\partial_jA_k$ and $E_i=\p_0 A_i-\p_i A_0$ with $\epsilon^{ijk}\equiv\epsilon^{0ijk}$.

Once again, we may note that if the anomaly vanishes, that is, for $C =0$,
 the Hamiltonian formulation (\ref{pitheta}-\ref{H-time}) reduces to the known Hamiltonian formulation for the perfect relativistic fluid \cite{schutz1970perfect, schutz1971hamiltonian,holm1984relativistic,1984-MarsdenRatiuWeinstein}. We notice that in this case the Hamiltonian depends on Clebsch potentials only through $\xi_i$. This feature is lost in the presence of the anomaly, i.e., when
 $C\neq 0$, although the equations of motion (\ref{dJC}-\ref{TuC}) still do not contain the Clebsch potentials explicitly. 

We shall comment on the
the meaning of this explicit dependence on $\theta$ in the following sections. Here we just point out that the coefficient of $E_k$ in
 the last term of (\ref{H-time}) may be interpreted as an intrinsic electric dipole moment of the fluid.

So far we have considered the background gauge field as space- and time-dependent. An interesting special case is when the
magnetic field is time-independent. It is then possible to choose a vector potential $A_i$ which is
independent of time as well. Then the last term of (\ref{H-time}) can be integrated by parts and the Hamiltonian  takes the form 
\bea
	H &=&\int \left[\rho\sqrt{\mu^2+(\xi_i-A_i)^2}-P(\mu)-A_0\rho\right] d^3 x  
 \nonumber \\
	&-& \frac{C}{6}\int A_0\left[2\,\xi_iB^i
	+\epsilon^{ijk}(\xi_i-A_i)\p_j\xi_k\right] d^3 x \,.
 \label{H-notime}
\eea 
In this case, the explicit dependence on $\theta$ has disappeared and the Clebsch potentials only appear in
the combination $\xi_i$.

In the next section we discuss the effect of the anomaly on the Poisson structure of the Hamiltonian formulation derived in this section.

%%%%%%%%%%%%%%%%%%%%%
\section{Poisson Brackets}
%%%%%%%%%%%%%%%%%%%%%

The variational principle (\ref{S-notation}) which is linear in time-derivatives immediately
provides us with the canonically conjugate pairs $\theta, \pi_\theta$ and $\beta, \pi_\beta$. The Poisson brackets of all fields follow then from the canonical ones for the above fields
\bea
	\{\theta,\pi_\theta'\} &=& \{\beta,\pi_\beta'\}=\delta(\boldsymbol x-\boldsymbol x') \,,
 \la{PBcan} 
\eea	
where we have listed only the non-vanishing Poisson brackets. Here and below we use a concise notation omitting the spatial arguments of the fields so that, e.g., $\beta$ means $\beta(\boldsymbol x)$, $\pi_\theta'$ means $\pi_\theta(\boldsymbol x')$ etc.

The hydrodynamic equations of motion (\ref{dJC}-\ref{TuC}) can be formulated as equations 
written entirely in terms of $\rho$ and $\xi_i$ without an explicit dependence on the Clebsch parameters. Therefore, we shall  look for the possible Hamitlonian reduction
of (\ref{H-time},\ref{PBcan}).
The reduction consists of the dynamic reduction, i.e., the Hamiltonian should be expressible only in terms of the density $\rho$ and dynamic velocity $\xi_i$, and  the kinematic reduction, i.e., the closure of Poisson brackets of $\rho$ and $\xi_i$ without the use of the Clebsch parameters \cite{1967-SeligerWhitham}.

As we remarked before, with the inclusion of the anomaly,
the dynamic reduction is only partially successful. Namely, the Hamiltonian (\ref{H-time}) does depend on $\p_i\theta$ in the case of general time-dependent gauge field background. In the case of time-independent background the dynamic reduction is complete and the Hamiltonian (\ref{H-notime}) depends on the Clebsch parameters only through $\xi_i$.

Remarkably, the Poisson algebra of $\rho$ and $\xi_i$ is closed for any gauge field background so that the kinematic reduction is achieved. Indeed, after some straightforward calculations, we derive from (\ref{PBcan}) and the definition (\ref{xiClebsch}) the following set of Poisson brackets closed with respect to the fields $\rho$ and ${\xi}_i$,
\bea
 	\left\{\rho_+,\rho_+'\right\} &=& \frac{C}{3}\,B^i\partial_i\,\delta(\boldsymbol x-\boldsymbol x')\,, 
 \label{rhorho'} \\ 
 	\left\{\widetilde\xi_i,\rho_+'\right\} 
	&=&\p_i\,\delta(\boldsymbol x-\boldsymbol x')\,,
 \la{xirho'} \\
	\left\{\widetilde\xi_i,\widetilde\xi_j'\right\} 
	&=&  -\frac{\p_i\widetilde\xi_j-\p_j\widetilde\xi_i-\epsilon_{jik}B^k}{\rho_-}\,
	\delta(\boldsymbol x-\boldsymbol x') \,. 
 \la{xixi'}   
\eea
Here, for the sake of brevity, we introduced the following compact notation,
\bea
	\widetilde\xi_i &\equiv& \xi_i-A_i\,,
 \\
	\rho_\pm &\equiv& \rho\pm\frac{C}{6}\,\widetilde\xi_i\,B^i\,.
 \la{rhopm}
\eea
A comment on the first of these equations, namely, (\ref{rhorho'}), is appropriate at this
point. It is well known that the commutator of charge densities
is modified by a Schwinger term in the presence of an anomaly
for the corresponding symmetry \cite{anom-commutator1}.
This can be shown by explicit computation of the corrections to commutators via Feynman diagrams, the triangle diagram leading to the specific form
given.\footnote{The computation of modified commutators follows a procedure known as the
Bjorken-Johnson-Low method where correlators of currents at slightly unequal times are calculated
and
 a suitable equal-time limit is taken.}
It can also be seen from a 2-cocycle constructed in terms of the descent equations
which lead to the anomalies \cite{anom-commutator2}.
Our action effectively reproduces this in the Poisson brackets.
We may also note that an expression analogous to (\ref{rhorho'}) has appeared in \cite{1998-AlekseevCheianovFrohlich}. 

We remark here that the dynamic velocity $\widetilde{{\xi}_i}$ and the modified densities $\rho_\pm$ are invariant under the transformations (\ref{gt}), therefore, the Poisson algebra (\ref{rhorho'}-\ref{xixi'}) is written in terms of explicitly gauge-invariant quantities. However, as it is well known \cite{anom-commutator2} generator of gauge transformations cannot be realized canonically in the presence of anomaly (see Sec.~\ref{sec:symgen}).

The algebra (\ref{rhorho'}-\ref{xixi'}) is obtained as a result of Hamiltonian reduction and is degenerate. It admits two Casimirs --- the quantities having vanishing Poisson brackets with fields entering Poisson algebra. They are given by
\bea
	C_1 &=& \int \rho_+\,d^3x\,,
 \la{C1} \\
	C_2 &=& \int \epsilon^{ijk}\,\widetilde\xi_i\,\partial_j\left(\widetilde\xi_k
	+2A_k\right) d^3x\,.
 \la{C2}
\eea
The charge density $j^0$ defined in (\ref{JuC}) is given by
\be
	j^0 = \rho_+ +\frac{C}{6}\epsilon^{ijk}\,\widetilde\xi_i\,\partial_j\left(\widetilde\xi_k
	+2A_k\right)\,.
 \la{j0}
\ee	
It is a combination of densities of two Casimirs of the algebra.

In the absence of anomaly $C=0$, all expressions (\ref{H-time}, \ref{rhorho'}-\ref{j0}) become the known formulas for perfect fluid dynamics \cite{1997-ZakharovKuznetsov, 2004-JNPPreview}.
Even when the anomaly is present, i.e., $C \neq 0$, 
if we consider the case of the background gauge field being absent,
we obtain again the formulas of anomaly-free hydrodynamics with a single exception. Namely, the definition of the charge density (\ref{j0}) still differs from $\rho$ by the density of Casimir (\ref{C2}). The latter is known as the helicity of the hydrodynamic flow.

Having Hamiltonian and Poisson brackets one can obtain equations of motion for any quantity $Q$ as $\dot{Q}=\p Q/\p t +\{H,Q\}$, where $\p Q/\p t$ denotes the ``explicit'' time-derivative. In our case this explicit derivative acts only on the time varying external gauge field. The dynamical
 fields $\xi_i$ and $\rho$ do not depend on time explicitly. For example, the equation of motion for $\xi_i$ will read
$\dot{\widetilde\xi}_i = -\p_t A_i +\{H,\widetilde\xi_i\}$, etc.

While the Clebsch variables appear in the 
algebra (\ref{rhorho'}-\ref{xixi'}) only via $\xi_i$, we should note that,
in the presence of the time-dependent gauge field background,
 the Hamiltonian (\ref{H-time}) contains $\p_i\theta$ in addition to the density and the dynamic velocity fields. Thus, the algebra (\ref{rhorho'}-\ref{xixi'}) is not adequate for a complete
 Hamiltonian description, and it should be supplemented by Poisson brackets involving
 the $\theta$ field. We list those brackets here for completeness
\bea
	\{\rho_+,\p_{k}\theta'\} &=& \p_{k}\delta(\boldsymbol x-\boldsymbol x')\,,
 \\
	\{\widetilde\xi_i,\p_{k}\theta'\} &=& \frac{\widetilde\xi_{i}+A_i-\p_i\theta}{\rho_-}\p_{k}\delta(\boldsymbol x-\boldsymbol x')\,.
\eea

%%%%%%%%%%%%%%%%%%%%%
\section{Symmetry generators}
\la{sec:symgen}
%%%%%%%%%%%%%%%%%%%%%

The Poisson algebra (\ref{rhorho'}-\ref{xixi'}) is closed and, in the case of the time-independent background, produces the hydrodynamic equations with the use of the Hamiltonian (\ref{H-notime}). However, the brackets (\ref{rhorho'}-\ref{xixi'}) are nonlinear and therefore do not have the Lie-Poisson form. For the symmetry analysis it is preferable to find an equivalent set of Poisson brackets corresponding to the algebra of symmetry generators of the system.

It is easy to see from (\ref{S-notation}) that the momentum densities can be defined as:
\be
	\Theta_{0i}=-\pi_\theta\partial_i\theta-\pi_\beta\partial_i\beta\,. 
 \la{Theta0i}
\ee
The momentum densities $\Theta_{0i}$ satisfy the diffeomosphism algebra and act as local translations in the absence of background field. However, one cannot express (\ref{Theta0i}) only in terms of the density $\rho$ and the dynamic velocity in the background of nonvanishing magnetic field. More precisely,  the canonical energy-momentum tensor acquires an explicit $\theta$ dependence:
\be
	\Theta_{0i} = \left(\rho+\frac{C}{6}\,A_kB^k\right)\, \xi_i
	+\frac{C}{6}\,B^k(\xi_k\partial_i\theta-\xi_i\partial_k\theta)\,.
 \la{Theta0ixi}
\ee

Let us now turn to gauge transformations which can be viewed as shifts in the field $\theta$. The naive canonical gauge generator for this symmetry is $-\pi_\theta$. Using (\ref{pitheta},\ref{xiClebsch}) we can write it as 
\begin{eqnarray}
	-\pi_\theta &=& \rho+\frac{C}{6}\,(A_i+\xi_i -\partial_i\theta)\, B^i \,
 \la{gaugeGen}
\end{eqnarray}
and notice that it also depends explicitly on $\partial_i\theta$.

It is straightforward to check that the Poisson structure (\ref{rhorho'}-\ref{xixi'}) can be put in a semidirect product Lie-Poisson algebra \cite{holm1984relativistic, 1984-MarsdenRatiuWeinstein} in terms of (\ref{Theta0ixi}, \ref{gaugeGen}). 

A gauge transformation of an arbitrary functional $F$ of basic fields generated by $-\pi_{\theta}$ is given by:
\be
	\delta_\Lambda F\equiv\int\left(-\Lambda(\boldsymbol{x}')\{\pi_\theta',F\}
	+\frac{\delta F}{\delta A_i (\boldsymbol x')}\p'_i\Lambda\right)d^3x'\,,
 \la{gtrans}
\ee
where the transformation of the gauge potential has also been added.

However, it is easy to see that (\ref{gtrans}) gives $\delta_\Lambda\alpha\neq0$, as well as $\delta_\Lambda\rho\neq0$ in apparent contradiction with their gauge invariance. In fact, one can show that the gauge symmetry (\ref{gt}) is not canonically realizable.

Let us now consider $\rho_{+}$ given by (\ref{rhopm}) as a generator of gauge transformations instead of $-\pi_{\theta}$. 
We easily check that  $\delta_\Lambda\alpha=\delta_\Lambda\beta\equiv0$ and $\delta_\Lambda\theta\equiv\Lambda$. 
Moreover, under the modified gauge transformations generated by $\rho_+$ the density $\rho$ transforms as:
\be
	\delta_\Lambda\rho=-\frac{C}{6}B^i\p_i\Lambda\,,
\ee
and there exists the gauge invariant quantity $\rho+\frac{C}{6}B^iA_i$. 

While $\rho_{+}$ can be considered as a generator of modified gauge transformations two subsequent gauge transformations generated by $\rho_{+}$ do not commute and the commutative algebra of gauge transformations has acquired a central extension (\ref{rhorho'}). This is, of course, a classical manifestation of a well known phenomenon in studies of quantum anomalies \cite{anom-commutator2}. At this point it is not clear whether similar modifications can be made for diffeomorphism generators (\ref{Theta0i})\footnote{The variables $\alpha$ and $\rho$ do not transform nicely under these generators.}. 

%%%%%%%%%%%%%%%%%%%%%
\section{Conclusion and discussion}
%%%%%%%%%%%%%%%%%%%%%

We have presented a variational principle for hydrodynamic equations with gauge anomaly at zero temperature. From the obtained action, we derived the Poisson structure and the Hamiltonian for the system. The most noteworthy feature of the obtained Hamiltonian formulation is that in the presence of gauge anomaly, the Hamiltonian reduction to the density and velocity fields is not complete and one of the Clebsch potentials becomes physical and is present in the Hamiltonian in the presence of the time-dependent gauge field background. 

The case of the time-independent external gauge fields is more natural for the Hamiltonian formulation. In this case one has a complete Hamiltonian reduction with both Hamiltonian and Poisson brackets expressed purely in terms of the charge density $\rho$ and dynamic velocity ${\xi}_i$.     

It turns out, however, that the generators of gauge transformations $\rho_{+}$ (\ref{rhopm}) cease to commute and that the generators of spatial translations (\ref{Theta0ixi}) can be written only with the explicit use the Clebsch potential $\theta$. 
The origin of the explicit appearance of $\theta$ in the Hamiltonian and in
(\ref{gaugeGen}) and ({\ref{Theta0i}) can be traced to the term $A \wedge \xi \wedge d \xi =
A \wedge d\theta \wedge d\alpha \wedge d\beta$
in the action. This term is needed in the hydrodynamic action to make sure that the anomalous charge non-conservation corresponds to the one obtained from the computation of the triangle diagram in the underlying QFT. In our variational approach the presence of $A \wedge \xi \wedge d \xi$ term does not lead to any entropy production and is in agreement with the requirement of positive semidefinite entropy production which was central in Son and Surowka analysis \cite{2009-SonSurowka}. A connection of the anomalous term with the entropy arguments might become more explicit if the variational principle could be generalized to finite temperature hydrodynamics. A possibility of such a generalization is worthy of further investigation.

We would like to acknowledge the Simons Center for Geometry and Physics for the support of the program on quantum anomalies in hydrodynamics in Spring 2014 during which this work has been done. We are grateful to V. Cheianov and B. Khesin and other participants of that program for multiple discussions and suggestions related to this work. 
The work of A.G.A. was supported by the NSF under grant no. DMR-1206790 and the work of V.P.N. was supported by NSF grant number PHY-1213380 
and by PSC-CUNY awards.

%\bibliographystyle{my-refs}

%\bibliographystyle{plain}
%\bibliographystyle{unsrt}
%\bibliographystyle{iopart-num}
%\bibliographystyle{abbrv}
%\bibliographystyle{apsrev}
%\bibliographystyle{amsplain}
%\bibliographystyle{ieeetr}
%\bibliographystyle{apsrev4-1}
%\bibliographystyle{aipnum4-1.bst}

%\bibliographystyle{unsrtnat}

%%%%%%%%%%%%%%%%%%%%%%%%%%%%%%%%%%%%%%%%%%%%%%
%%%%%%%%%%%%%%%%%%%%%%%%%%%%%%%%%%%%%%%%%%%%%%
%%%%%%%%%%%%%%%%%%%%%%%%%%%%%%%%%%%%%%%%%%%%%%
%%%%%%%%%%%%%%%%%%%%%%%%%%%%%%%%%%%%%%%%%%%%%%
\end{document}